\documentclass{emulateapj}

\shortauthors{Alvarez, Bromm, \& Shapiro}
\shorttitle{HII Region of the First Star}

\makeatletter

\newenvironment{inlinefigure}{%
\def\@captype{figure}%
\noindent\begin{minipage}{0.999\linewidth}\begin{center}}
{\end{center}\end{minipage}}
\makeatother
 
\def\ltsima{$\; \buildrel < \over \sim \;$}
\def\simlt{\lower.5ex\hbox{\ltsima}}
\def\gtsima{$\; \buildrel > \over \sim \;$}
\def\simgt{\lower.5ex\hbox{\gtsima}}

\newcommand{\beq}{\begin{equation}}
\newcommand{\eeq}{\end{equation}}

\begin{document}

\title{The \ion{H}{2} Region of the First Star}

\author{Marcelo A. Alvarez, Volker Bromm, and Paul R. Shapiro}
\affil{Department of Astronomy, University of Texas at Austin, Austin, TX 
78712;\\
marcelo@astro.as.utexas.edu,
vbromm@astro.as.utexas.edu, 
shapiro@astro.as.utexas.edu
}

\begin{abstract}

Simulations predict that the first stars in a $\Lambda$CDM universe
formed at redshifts $z>20$ in minihalos with masses of about $10^6
M_{\sun}$.  We have studied their radiative feedback by simulating the
propagation of ionization fronts (I-fronts) created by these first
Population~III stars ($M_* = 15-500 M_{\sun}$) at $z=20$, within the
density  field of a cosmological simulation of primordial star
formation, outward thru the host minihalo and into the surrounding
gas.  A three-dimensional ray-tracing calculation tracks the I-front
once the \ion{H}{2} region evolves a ``champagne flow'' inside the
minihalo,  after the early D-type I-front detaches from the shock and
runs ahead, becoming R-type.   We take account of the hydrodynamical
back-reaction by an approximate model of the central wind.  We find
that the escape fraction of ionizing radiation from the host halo
increases with stellar mass, with $0.7\la f_{\rm esc} \la 0.9$ for
$80\la M_*/M_\odot \la 500$.  To quantify the ionizing efficiency of
these stars as they begin cosmic reionization, we find that, for
$M_*\ga 80M_\odot$, the ratio of gas mass ionized to stellar mass  is
$\sim 60,000$, roughly half the number of ionizing photons released
per stellar baryon.  Nearby minihalos are shown to trap the I-front,
so their centers remain neutral.  This is contrary to the recent
suggestion that these stars would trigger formation of a second
generation by fully ionizing neighboring minihalos, stimulating H$_2$
formation in their cores.  Finally, we discuss how the evacuation of
gas from the host halo reduces the growth and luminosity of
``miniquasars'' that may form from black hole remnants of the first
stars.

\end{abstract}


\section{Introduction}

The formation of the first stars marks the crucial transition from an
initially simple, homogeneous universe to a highly structured one at
the end of the cosmic ``dark ages'' (e.g., Barkana \& Loeb 2001; Bromm
\& Larson 2004; Ciardi \& Ferrara 2005).  These so-called
Population~III (Pop~III) stars are predicted to have formed in
minihalos with virial temperatures  $T\lesssim 10^4$ K at redshifts
$z\gtrsim 15$ (e.g., Couchman \& Rees 1986; Haiman, Thoul, \& Loeb
1996; Gnedin \& Ostriker 1997; Tegmark et al. 1997; Yoshida et
al. 2003). Numerical simulations are indicating that the first stars,
forming in primordial minihalos, were predominantly very massive stars
with typical masses $M_{\ast}\ga 100 M_{\odot}$ (e.g., Bromm, Coppi,
\& Larson 1999, 2002; Nakamura \& Umemura 2001; Abel, Bryan, \& Norman
2002).  In this paper, we investigate the question: {\it How did the
radiation from the first stars ionize the surrounding medium,
modifying the conditions for subsequent structure formation?} This
radiative feedback from the first stars could have
played an important role in the reionization of the universe (e.g.,
Cen 2003; Ciardi, Ferrara, \&  White 2003; Wyithe \& Loeb 2003;
Sokasian et al. 2004).

Recent observations of the large-angle polarization anisotropy of the
cosmic microwave background (CMB) with the Wilkinson Microwave
Anisotropy  Probe (WMAP; Spergel et al. 2003)  imply a free electron
Thomson scattering optical depth of $\tau=0.17$, suggesting that the
universe was substantially ionized by a redshift $z=17$ (Kogut et
al. 2003).  Such an early episode of reionization may require a
contribution from massive Pop~III stars  (e.g., Cen 2003; Wyithe \&
Loeb 2003; Furlanetto \& Loeb 2005).  Analytical and numerical studies
of reionization typically parametrize the efficiency with which these
stars reionize the universe in terms of quantities such as the escape
fraction and fraction of baryons able to form stars (e.g., Haiman \&
Holder 2003).  In order to
understand the role of such massive stars in reionization, it is
therefore crucial to understand in detail the fate of the ionizing
photons they produce, taking proper account of the structure within
the host halo.

Until now, studies of the propagation of the ionization front
(I-front) within the host halo have been limited to analytical or
one-dimensional numerical calculations (Kitayama et al. 2004;
Whalen et al. 2004).  These studies suggest that the escape fraction
is nearly unity for small halos, and as shown by Kitayama et al.  (2004),
is likely to be much smaller for larger halos.  These conclusions
should not be taken too literally, however, since the structure of the
halos in which the stars form is inherently three-dimensional.
Rather, that work should be viewed as laying the foundation for more
detailed study in three dimensions.  An effort along these lines was
recently reported by O'Shea et al. (2005), focused on the dynamical
consequences of the relic \ion{H}{2} region left by the death of the
first stars.

Here, we present three-dimensional calculations of the
propagation of an I-front through
the host halo ($M\simeq 10^6 M_\odot$) and into the intergalactic
medium (IGM).  We model the hydrodynamic feedback that results from
photoionization heating through the use of the similarity solutions
developed by Shu et al. (2002; ``Shu solution''), and calculate the
propagation of the I-front by following its progress along
individual rays that emanate from the 
\begin{inlinefigure}
\resizebox{8cm}{!}{\includegraphics{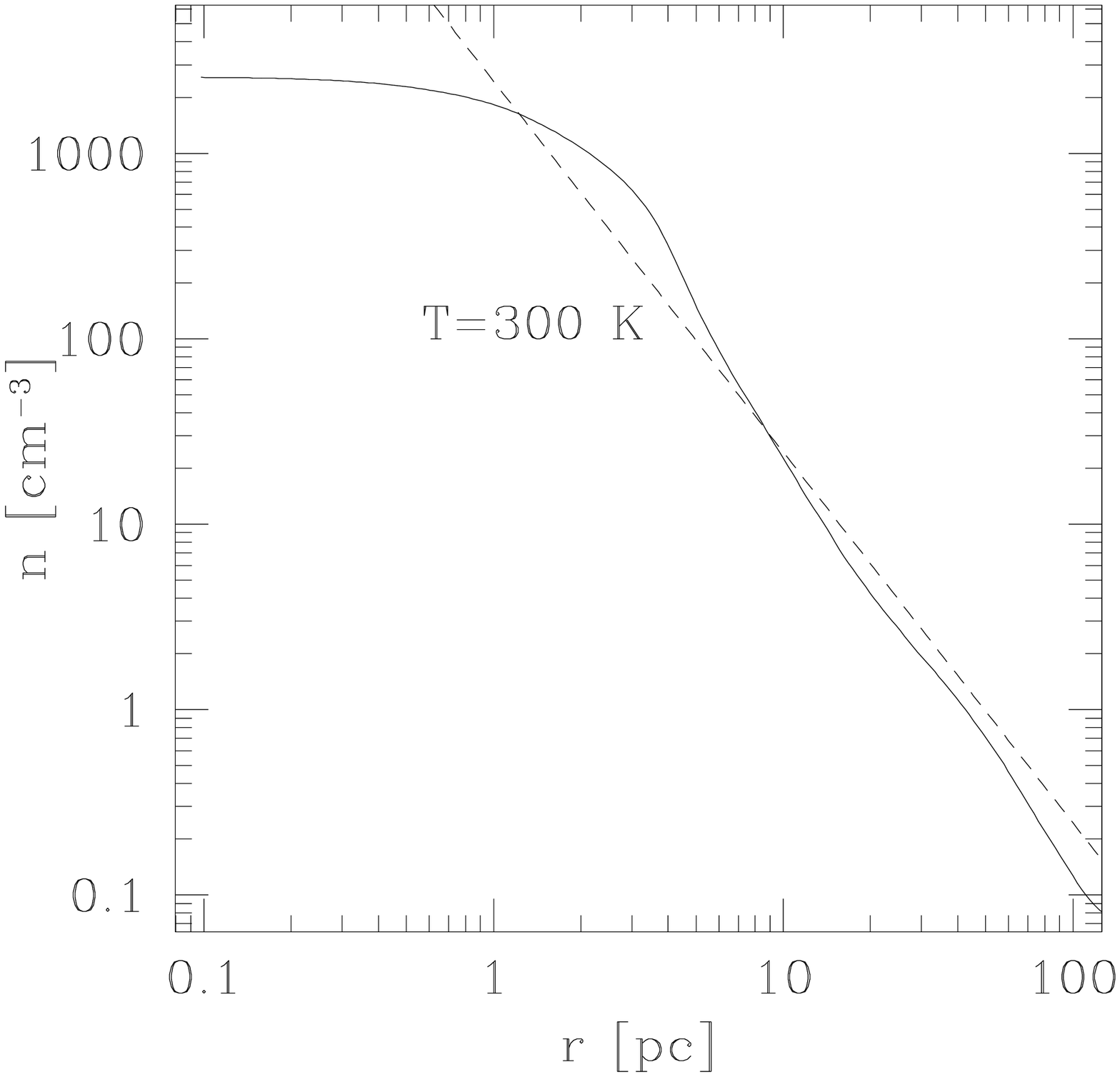}}
\caption{Hydrogen number density profile in a halo of mass $M=10^6
M_\odot$ at $z=20$. {\it Solid line:} Spherically-averaged
density profile of minihalo in the SPH simulation.
{\it Dashed line:} Density profile for a SIS with $T_{\rm SIS}=300$~K.
\smallskip
\label{plot1}
}
\end{inlinefigure}
star.

This paper is organized as follows.  In \S2 we present an analytical
estimate of when and if the I-front escapes the host halo and
describe our application of the self-similar Shu solution to
primordial star forming halos.  We describe  our numerical method in
\S3, and present our results in \S4.  Finally,  we discuss the
implications of our calculations  for reionization and  further star
formation in \S5.

\section{Physical Model for Time-dependent \ion{H}{2} Region}

At early times, as the I-front begins to propagate away from the star, 
its evolution is coupled to the hydrodynamics of the gas.  The effect
of this hydrodynamic response is to lower the density of gas as it
expands in a wind, and eventually the I-front breaks away from the 
expanding hydrodynamic flow, racing ahead of it.  In what follows, we 
describe an approximate, spherically-symmetric model for the
relation between the I-front and the hydrodynamic flow in the center
of the halo.  This
allows us to account for the consumption of ionizing photons
within the halo while at the same time tracking the three-dimensional 
evolution of the
I-front after it breaks out from the center of the halo and propagates 
into the surrounding IGM.

\subsection{Early Evolution}

In a static density field, it takes of order a recombination time for
an I-front propagating away from a source that turns on
instantaneously to slow to its ``Str\"omgren radius" $r_{\rm S}$, at which
point recombinations within balance photons being emitted by the
source.  Generally, the I-front moves supersonically until it approaches
the Str\"omgren radius, at which point it must become subsonic before
it slows to zero velocity.   The supersonic evolution of the I-front
is generally referred to as ``R-type" (rarefied), whereas the subsonic
phase is referred to as ``D-type"
(dense).  In the R-type phase, the I-front races ahead of the
hydrodynamic 
response of the photoheated gas.   In the D-type phase,
however, the gas is able to respond hydrodynamically, and a shock
forms ahead of the I-front (see, e.g., Spitzer 1978).

For the case considered here of a single, massive Pop~III
star forming in the center of a minihalo, this initial R-type phase
when the star first begins to shine is likely to be very short lived,
of order the recombination time in the star-forming cloud, $t_{\rm
rec}<1$ yr for $n\simeq 10^6$ cm$^{-3}$.  This time is even shorter
than the time it takes for the star to reach the main sequence,
$t\simeq 10^5$ yr, given by the Kelvin-Helmholtz time.  Thus,
hydrodynamic effects are likely to be dominant at early times when the
I-front is very near to the star.

The study of the formation of these stars at very small scales defines
the  current frontier of our understanding, where the initial gas
distribution and its interaction with the radiation emitted by the
star is highly uncertain (e.g., Omukai \& Palla 2001, 2003; Bromm \& Loeb 2004).
For example, it is still not yet known
whether a centrifugally supported disk  will form (e.g., Tan \& McKee
2004), or whether hydrodynamic processes can  efficiently transport
angular momentum outward, leaving a sub-Keplerian core (e.g., Abel
et al. 2002). Omukai \& Inutsuka (2002) studied the problem
in spherical  symmetry, making certain assumptions about the accretion
flow and the size of a spherical \ion{H}{2} region.  Since the
density and ionization structure in the immediate vicinity of
accreting Pop~III protostars is only poorly known, we
must also make some assumptions here about the progress of the
I-front at distances unresolved in the simulation we use ($\la 1$ pc).

We will therefore assume in all that follows that the hydrodynamic
response of the gas in the subsequent D-type phase will be to create a
nearly uniform density, spherically-symmetric wind, bounded by a
D-type I-front that is led by a shock.  The degree to which the gas
within this spherical shock is  itself spherically-symmetric depends
on the effectiveness with which the high interior 
pressure can reduce density inhomogeneities within.  
The timescale for this effect is the
sound-crossing timescale, comparable to the expansion timescale of the
weakly-supersonic D-type shock that moves at only a few times the
sound speed.  Since these two timescales are comparable, it is
plausible that pressure will be able to homogenize the density
structure behind the shock.  Furthermore, the one-dimensional
spherically-symmetric calculations of Whalen et al. (2004) and
Kitayama et al. (2004) indicate that pressure is capable of
homogenizing the density in {\em radius}, as shown by the nearly flat
density profiles found behind the shock in the D-type phase.  Although
these are reasonable assumptions for these first calculations, we
caution the reader that the detailed evolution of the I-front,
especially very close to the star itself, can only be
thoroughly understood by fully-coupled radiative transfer and
hydrodynamic simulations that resolve the accretion flow around the
star, which we defer to future study.

\begin{inlinefigure}
\resizebox{8cm}{!}{\includegraphics{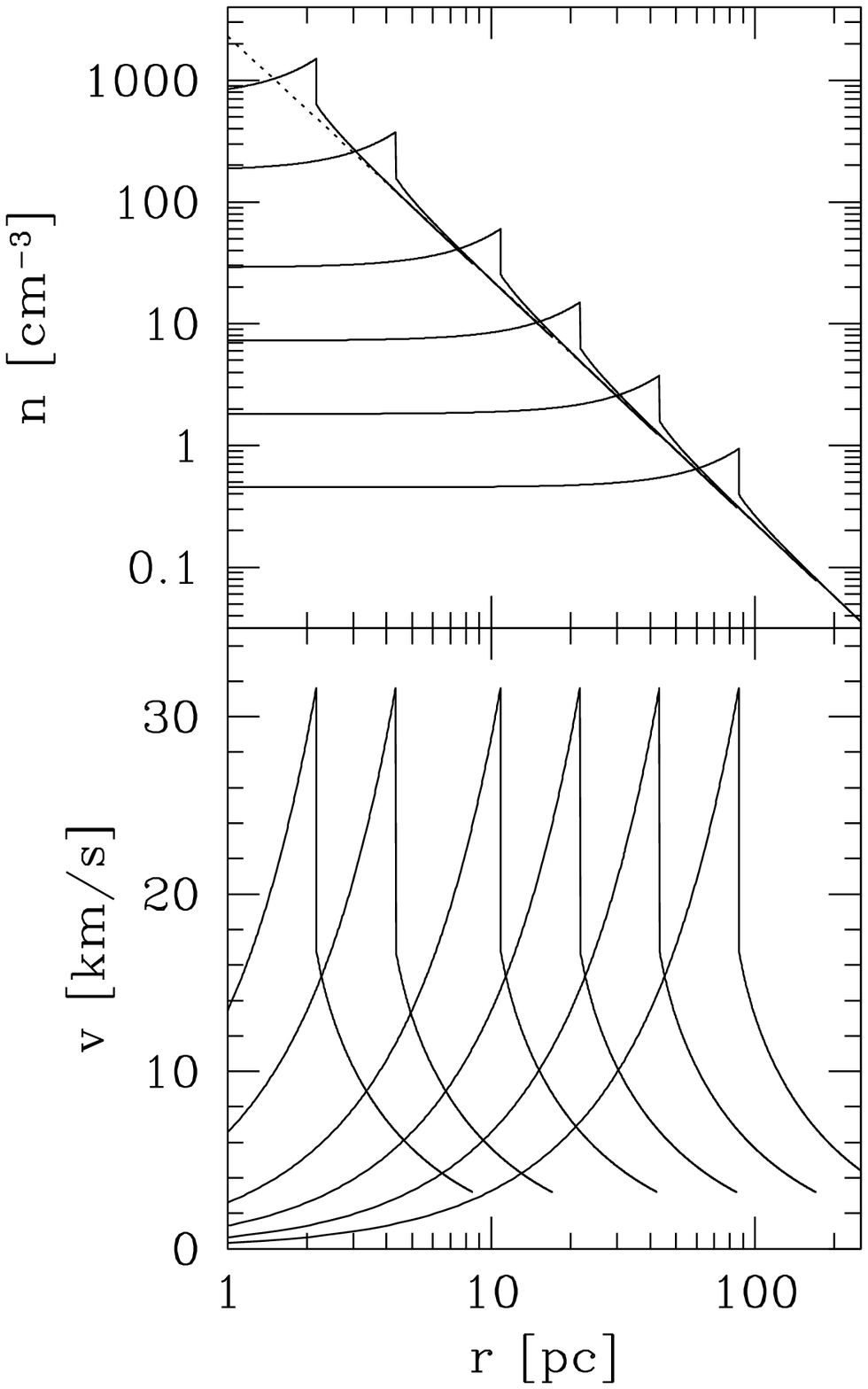}}
\caption{{\it Top:} Density profile given by the Shu solution at different times
after source turn-on,
$t=5\times 10^4$ ,$10^5$, $2.5\times 10^5$, $5\times 10^5$, 
$10^6$, and $2\times 10^6$ yr, from left to right.  {\it Bottom:} Same
as top but for velocity.  The peak velocity, in the post-shock gas
just behind the shock is constant in time, and is about 25\% lower
than the velocity of the shock itself.
\label{shuprofs}
}
\end{inlinefigure}

\subsection{Model for Breakout}

Primordial stars are expected to form enshrouded in a highly
concentrated distribution of gas.  For a star forming within a halo
with mass $\simeq 10^6 M_\odot$, the spherically-averaged density
profile of the gas, just prior to the onset of protostellar collapse,
is well-approximated  by that of a singular
isothermal  sphere (SIS) with a temperature $\sim 300$ K (Figure 1).
For values relevant to a star-forming region in a minihalo with mass
$M\sim 10^6 M_\odot$ at $z\sim 20$, the hydrogen atom number density
profile of the SIS is given by  
\beq 
n(r)\simeq 2.3\times 10^3
\left(\frac{T_{\rm SIS}}{300{\rm K}}\right) \left(\frac{r}{1{\rm
pc}}\right)^{-2} {\rm cm^{-3}},
\label{sis}
\eeq  
where we have assumed a hydrogen mass fraction $X=0.75$.  We
will  take the SIS as a fiducial density profile for the calculations
presented here.

As discussed in \S2.1, a D-type shock initially propagates outward
just ahead of the I-front,  leading to an outflow and
corresponding drop in central density. After some time $t=t_B$,
however, the central density is sufficiently lowered so that
recombinations can no longer trap the I-front behind the
shock, and it quickly races ahead.

This ``breakout'' time $t_B$ marks the moment at which ionizing
radiation is  no longer bottled up within and can escape.   If the
lifetime of the star $t_* < t_B$, then  the front never escapes and
the escape fraction is zero.  This is essentially the reasoning used
by Kitayama et al. (2004) to explain their result that  ionizing
radiation does not escape from halos with mass $M > M_{\rm crit}$,
where $M_{\rm  crit}$ is determined by setting $t_*=t_B$.

After breakout, the gas left behind is close to isothermal with the high
temperature of a photoionized gas (a few times $10^4$ K).
The strong density gradient
results in a pressure imbalance that drives a wind outward, bounded by
an isothermal shock.  This ``champagne''  flow (e.g.,  Franco et
al. 1990) has been analyzed through similarity  methods by Shu et
al. (2002), who found self-similar solutions for different  power-law
density  stratifications $\rho\propto r^{-n}$, $3/2<n<3$, and is also
evident in the one-dimensional calculations of Whalen et al. (2004)
and Kitayama et al. (2004).   The family of solutions obtained by Shu
et al. (2002) for the $n=2$ case are  described in terms of the
similarity variables (eqs. (12) and (13) of Shu et al. 2002)
\beq x=\frac{r}{c_st}
\label{ndimx}
\eeq
and
\beq \rho(r,t)=\frac{m_pn(r)}{X}=\frac{\alpha(x)}{4\pi Gt^2},
\label{ndimrho}
\eeq
where $c_s$ is the sound speed of the ionized gas and $\alpha(x)$
characterizes the shape of the density profile in the champagne flow.
If gas within the initial SIS
has a  sound speed $c_{s,1}$, then different solutions are obtained for
$\alpha(x)$,  depending on the ratio $\epsilon\equiv
(c_{s,1}/c_s)^2$, where $c_{s,1}$ and  $c_s$ are the initial SIS
sound speed and ionized gas sound speed,  respectively.  For $T_1\sim
300K$ and $T\sim 2\times 10^4K$,  $\epsilon\sim 0.007$ and the shock
moves at $v_s=x_sc_s\simeq 40$ km s$^{-1}$, where $x_s=2.55$.  
In Figure \ref{shuprofs}, we have plotted the density and velocity profiles
in the Shu solution for the above parameters.  As seen in the figure,
the density drops steadily in the center and is nearly uniform, while the
velocity profile is unchanged as it moves outward.  The peak velocity,
corresponding to post-shock gas, is constant in time, $\sim 30$ km s$^{-1}$,
and is less than the velocity of the shock itself.

In what
follows, we describe a
model for when and where breakout occurs by finding the moment in the
post-breakout Shu solution where the recombination rate inside of the
shock is equal to the ionizing luminosity of the star.
The condition for breakout is
\beq Q_*=4\pi\alpha_B\int_0^{r_{\rm sh}(t_B)}r^2 n^2(r,t_B)dr,
\label{dimcrit}
\eeq
where $\alpha_B=1.8\times 10^{-13}\ {\rm cm}^3 \ {\rm s}^{-1}$ is the
recombination rate coefficient to excited states of hydrogen at  $T\sim
2\times10^4{\rm K}$, $n(r,t)$ is the number density in the Shu
solution, $Q_*$ is the ionizing photon luminosity of the star,  and
$r_B\equiv r_{\rm sh}(t_B)=c_sx_st_B$ is the position of the  shock at
breakout.  As shown by Bromm, Kudritzki, \& Loeb (2001),  the ionizing
luminosity of primordial stars with masses $M>100 M_\odot$  is
roughly proportional to the mass of the star, $Q_* \simeq 1.5\times10^{50}$
s$^{-1} (M/100M_\odot)$ (see Schaerer 2002 for more detailed
calculations).

Combining equations (\ref{ndimx}), (\ref{ndimrho}), and
(\ref{dimcrit}), we can solve for the breakout radius
\beq r_B=\frac{\alpha_B c_s^4x_S}{4\pi (\mu m_pG)^2Q_*}\int_0^{x_S}
\alpha^2(x)x^2dx.  \eeq
For fiducial values, we obtain 
\beq 
r_B\simeq 2.3 {\rm pc} \left(\frac{T_{\rm SIS}}{300 {\rm K}}\right)^2 \left(\frac{Q_*}{3\times 10^{50}s^{-1}}\right)^{-1}.  
\eeq 
Parameterizing the speed of the shock as
$v_s$, we can use the  formula $t_B=r_B/v_s$ to derive the time after
turn on at which breakout occurs, 
\beq 
t_B \simeq 5.6\times 10^4 {\rm yr} 
\left(\frac{v_s}{40\ {\rm km/s}}\right)^{-1} \left(\frac{T}{300
{\rm K}}\right)^2 \left(\frac{Q_*}{3\times 10^{50}
s^{-1}}\right)^{-1}.  
\eeq 
Here and in the calculations we will present, we assume that the speed
of the shock front in the D-type and champagne phases is the same, 
$v_s\simeq 40$ km s$^{-1}$.  The lifetimes of massive stars with masses
$100<M/M_\odot<500$ are  within the range $2\la t_*\la 3$~Myr
(e.g., Bond, Arnett, \& Carr 1984), much longer
than our estimate of  $t_B\sim 5.6\times 10^{4}$ yr for our fiducial
values.  Thus, we expect the time-dependent fraction of 
ionizing photons that escape from the halo to rapidly approach unity
over this mass range.

For lower stellar masses, and therefore lower values of $Q_*$, the breakout
time $t_B$ becomes comparable to the stellar lifetime $t_*$, which
itself increases with decreasing mass (see Figure \ref{tvm}).  The precise value of the
stellar mass at which $t_B=t_*$ is therefore quite sensitive to the
speed of the D-type shock, the density profile used in equation 
(\ref{dimcrit}), and, of course, the main sequence lifetime of the star.
In particular, the early hydrodynamic behavior of the gas in the
D-type phase depends on an extrapolation to small scales where the
mass distribution is not well understood.  While our model for
breakout is consistent with the one-dimensional calculations of 
Whalen et al. (2004) and Kitayama et al. (2004) in predicting that
the escape fraction for massive stars $M_* > 100M_\odot$ approaches
unity because breakout occurs early in their lifetimes,
the threshold stellar mass at which $t_B=t_*$ and the escape
fraction goes to zero is not well determined and deserves further attention.

\section{Numerical Methodology}
\subsection{Cosmological SPH Simulation}
The basis for our radiative transfer calculations is a 
cosmological simulation of high-$z$ structure formation
that evolves both the dark matter and baryonic components,
initialized according to the $\Lambda$CDM model at $z=100$,
to $z=20$. We use the GADGET code that combines a tree, hierarchical
gravity solver with the smoothed particle hydrodynamics (SPH) method
(Springel, Yoshida, \& White 2001). In carrying out the cosmological
simulation used in this study, we adopt the same parameters as
in earlier work (Bromm, Yoshida, \& Hernquist 2003). Our periodic box size,
however, is now
$L=200 h^{-1}$ kpc  comoving. Employing the same number of 
particles, $N_{\rm DM}=N_{\rm SPH}=128^3$, as in Bromm et al. (2003), the SPH
particle mass here is $\sim 70 M_\odot$.  

\begin{inlinefigure}
\resizebox{8cm}{!}{\includegraphics{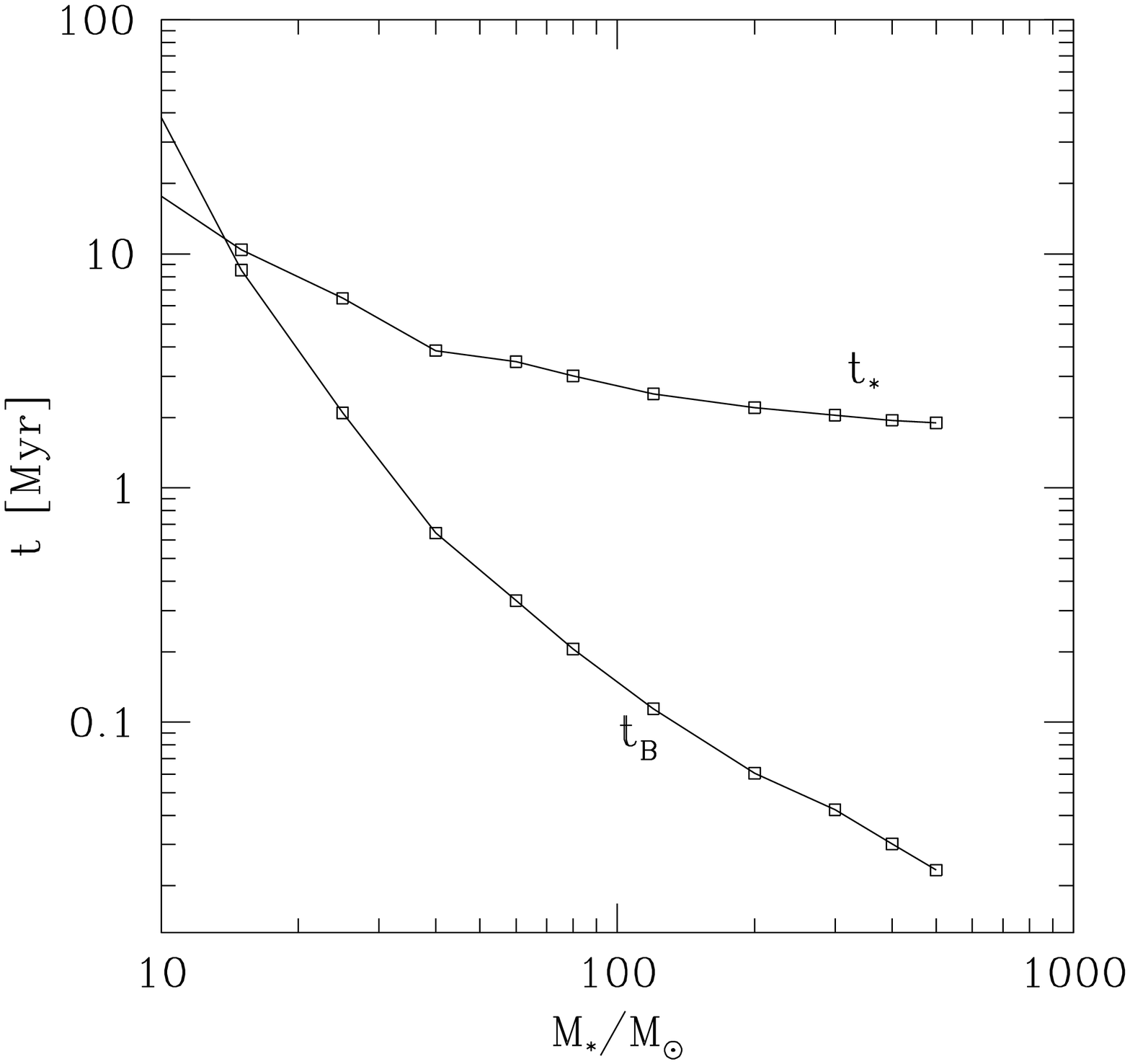}}
\caption{ Timescales for breakout $t_B$ and stellar lifetimes
$t_*$ versus stellar mass $M_*$. Our estimate for $t_B$ becomes
increasingly uncertain toward smaller stellar mass. Within these
uncertainties, we estimate that no ionizing radiation will escape
into the IGM for $M_* \la 15 M_{\odot}$.
\smallskip
\label{tvm}
}
\end{inlinefigure}
We place the point source, 
representing the already fully formed Pop~III star, at the
location of the highest density SPH particle in the simulation at
$z=20$, $n\sim 10^4$ cm$^{-3}$, located within a halo of mass $M_{\rm
vir}\sim 10^{6} M_\odot$  and virial radius $r_{\rm vir}\sim 150$ pc.

\subsection{Ray Casting}

To calculate the propagation of the I-front, we construct a
set of $N_R=12\times 2^{12}=49152$ rays around the source.  The
directions  of each of the rays are chosen to lie in the center of a
HEALPixel\footnote{http://www.eso.org/science/healpix}  (G\'orski et
al. 2005).  Each ray represents an equal solid angle of the sky as
seen from the source. The rays are discretized into $N_s=640$ segments
of length $\Delta r_i$ each, logarithmically spaced in radius, so that
the radial  coordinate of the front of the $i$th segment is $r_i$.
The inner radius is $\simeq 0.2$ pc (proper), while the outer radius
$\simeq 12$ kpc (proper). In the case of a single source,  we treat
the evolution along each ray as independent of every other ray,  so
that we may represent each ray as a different spherically symmetric
density field.

In order to calculate the evolution of the I-front, we must
know the  density of hydrogen along the ray.  We do this by
means of interpolation from a mesh upon which the density is
precalculated.  The density within each segment is assigned from the
mesh at the midpoint of the ray segment,
\beq  r^3_{i+1/2}\equiv\frac{1}{2}\left(r^3_i+r^3_{i+1}\right).  \eeq
This discretization ensures that the midpoint of the ray is located at
the point at which half the mass within the volume element is at
$r<r_{i+1/2}$ and the other half is at $r>r_{i+1/2}$.  The density
value at the midpoint is determined by tri-linear interpolaton from
the eight nearest nodes on the mesh,  
\beq 
n(x_m,y_m,z_m)=\sum_{g=1}^8n_g f(x_g)f(y_g)f(z_g),  
\eeq  
where $f(x_g)\equiv 1-|x_m-x_g|/\Delta_c$, 
$\Delta_c$ is the mesh cell
size, $(x_g,y_g,z_g)$ are the coordinates of the eight nearest grid 
points, $n_g$ is the density at grid point $g$,  
\begin{eqnarray}
x_m=&r_{i+1/2}\sin\theta\cos\phi, \\ y_m=&r_{i+1/2}\sin\theta\sin\phi,
\\ z_m=&r_{i+1/2}\cos\theta,
\end{eqnarray}
and ($\theta,\phi$) are the angular coordinates of the ray.

Given the substantial dynamic range necessary to resolve the \ion{H}{2}
region around a Pop III star, the  use of only one uniform mesh to
interpolate between the SPH density field and the rays is not
possible.  Here we make use of the fact that the system is highly
centrally concentrated, which allows  for the use of a set of
concentric equal-resolution uniform meshes, each one half the linear
size of the
last,  centered on the star forming region in the center of the halo.
Since the segments are spaced logarithmically in radius, the segment
size at any point is smaller than the mesh cell of the highest
resolution mesh that overlaps that point.  We interpolate the SPH
density to each of the hierarchical meshes (see Appendix).  For each
ray segment midpoint, we find the highest resolution mesh overlapping
that point and use tri-linear interpolation from that mesh, as
described above.

\subsection{Ionization Front Propagation}

In deriving the I-front evolution, we make the approximation  that the
front is sharp -- i.e. gas is completely ionized inside and completely
neutral outside.  Because the equilibration time is short on the
ionized side, every recombination is balanced by an absorption.  Under
this  assumption,  the I-front ``jump condition'' (Shapiro \&
Giroux 1987) implies a differential  equation for the evolution of the
I-front radius (Shapiro et al. 2005; Yu 2005),
\beq 
\frac{dR}{dt}=\frac{cQ(R,t)}{Q(R,t)+4\pi R^2cn(R)},
\label{diffeq}
\eeq
where $Q(R,t)$ is the ionizing photon luminosity at the surface of the
front.  This equation correctly takes into account the finite travel
time of ionizing photons (e.g., {$\dot{R}\rightarrow c$ as
$Q(R,t)\rightarrow \infty$).  In general,  this equation can be solved 
numerically, once $Q(R,t)$ and $n(R)$ are known.

To approximate the hydrodynamic response due to photoheating, we
combine  the Shu solution with our ray tracing method to follow the
I-front after it breaks out into the rest of the halo.  We
assume that the front makes an initial transition from R to D-type at
radii $r\ll 1 {\rm pc}$, and   creates a spherical D-type front that
propagates outward to the breakout  radius $r_B$, after which the Shu
solution is expected to be valid.  For  each ray, we assume that the
density profile of the gas at breakout is given by the self-similar
solution inside of the shock, and is undisturbed outside, given
by our cosmological SPH simulation (see \S3.1).

The initial I-front radius is independent of angle and is
initialized to the breakout radius, so that equation (\ref{diffeq}) is
solved with the initial value $R(t_B)=r_B$.  $Q(R,t)$ in 
\begin{inlinefigure}
\resizebox{8cm}{!}{\includegraphics{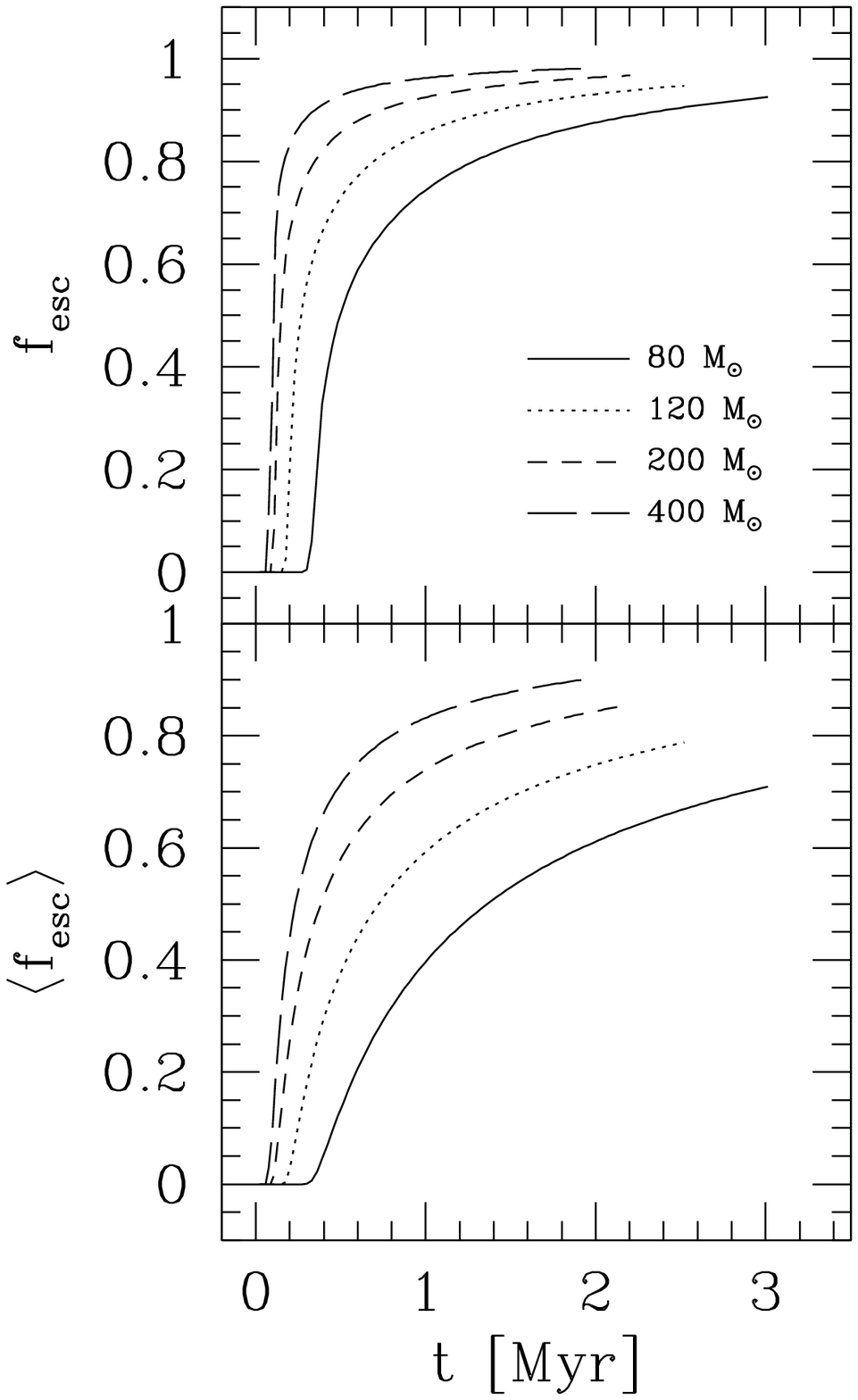}}
\caption{{\it Top:} Instantaneous escape fraction for different masses, 
as labeled.  
{\it Bottom:} Time-averaged escape fraction $\langle f_{\rm esc}\rangle$  as
defined in the text.  Although the instantaneous escape fraction rises
quickly just after breakout, the time-averaged value retains memory of 
the breakout time and therefore lags behind.
\smallskip
\label{fvt}}
\end{inlinefigure}
equation
(\ref{diffeq}) depends on the density profile along a ray, and is
given by
\beq Q(R,t)=Q_*-4\pi\alpha_B\int_0^R n^2(r,t)r^2dr, \eeq where
$n(r,t)$ is given by the Shu solution at $t$ for all  $r<r_{\rm
sh}(t)$, while for $r>r_{\rm sh}(t)$ the density is the  initial
unperturbed angle-dependent density distribution along each  ray.
We assume that the ionizing photon luminosity is 
constant over the lifetime of the star, with values given in Table 4
of Schaerer (2002). The final time in each run is set to the
corresponding stellar lifetime for each mass, also given in Table 4 of
Schaerer (2002).
\section{Results}

We have carried out several ray-tracing runs, each for a different stellar
mass forming within the same host minihalo.

\subsection{Escape Fraction}

The fraction of the ionizing photons emitted by the central
star which escape into the IGM beyond the virial
radius of the host minihalo is a fundamental ingredient in the theory
of cosmic reionization and of the feedback of Pop~III star formation
on subsequent star and galaxy formation.  We use our \ion{H}{2} region
calculations to 
\begin{inlinefigure}
\resizebox{8cm}{!}{\includegraphics{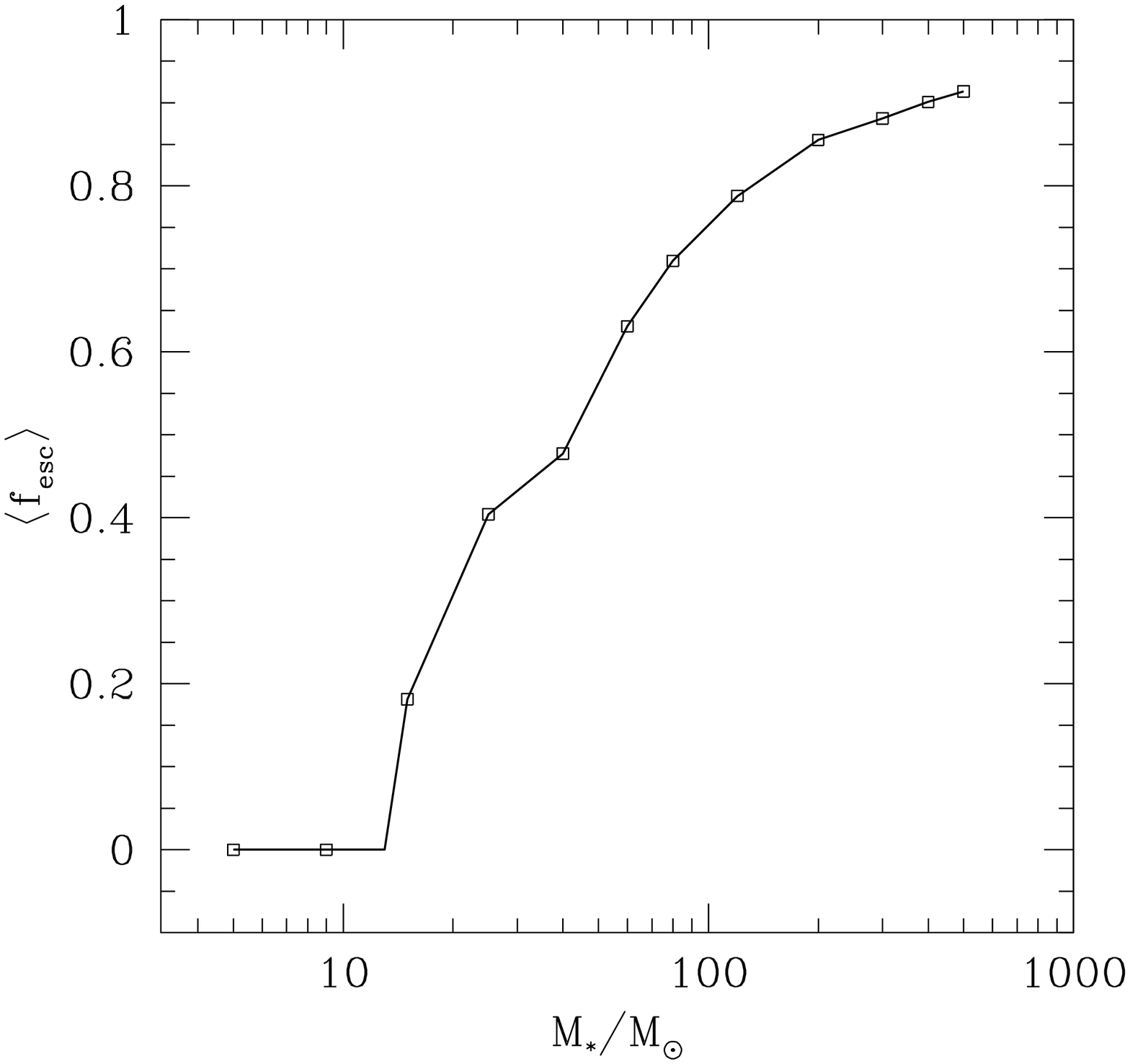}}
\caption{ Mean escape fraction at the
end of the star's lifetime $t_*$, versus stellar mass.  Each symbol
corresponds to $\langle f_{\rm esc} \rangle$ as defined by Eq. 
(\ref{fescmean}) for a different stellar mass calculation.
Note that the escape fraction approaches zero for $M\la 15 M_\odot$,
for which $t_B\ga t_*$. 
\label{fvm}
}
\end{inlinefigure}
derive this escape fraction $f_{\rm esc}$ and its
dependence on time during the lifetime of the star.  Since our \ion{H}{2}
region density field and radiative transfer are three-dimensional, the
escape fraction is angle-dependent. Along each ray, the escape
fraction is given by
\beq 
f_{\rm esc}(t) =\left\{
\begin{array}{ll}
1-\frac{4\pi\alpha_B}{Q_*}\int_0^{r_{\rm vir}}n^2(r,t)r^2dr, & 
R(t)>r_{\rm vir},\\
0, & R(t)\leq r_{\rm vir},
\end{array}
\right.
\eeq 
where $r_{\rm vir}=150$ pc and $n(r,t)$ is given by the Shu solution for
$r<r_{\rm sh}(t)$, and by the SPH density field in that direction for
$r>r_{\rm sh}(t)$.   The instantaneous escape fraction versus time, $f_{\rm
esc}$, determined
by taking the average over all angles of the angle-dependent escape
fraction
is shown in the top panel of Figure \ref{fvt}. The average escape
fraction between turn-on and time $t<t_*$ is given by  
\beq 
\langle f_{\rm esc}\rangle \equiv \frac{1}{t}\int_0^t f_{\rm esc}(t')dt',
\label{fescmean}
\eeq 
and is shown in the bottom panel of Figure \ref{fvt}.
Figure \ref{fvm} shows the average escape fraction at the end of the
star's lifetime versus mass.  For the very high mass  $M_*=500
M_\odot$ case, the mean escape fraction is $\langle f_{\rm esc}\rangle\sim
0.9$, while for $M_*=80 M_\odot$ it is about 0.7. We can understand
the zero lifetime-averaged escape fraction at the smallest masses, as evident in
Figure \ref{fvm}, by comparing
the lifetime of the star and the
breakout time.  For $t_B<t_*$ breakout occurs before the star dies and
the escape fraction is expected to be greater than zero.  For
$t_B>t_*$, little or no radiation should escape.  As can be seen in
Figure \ref{tvm}, the threshold mass for which $t_B=t_*$ is about 15 $M_\odot$.
However, as we discussed in \S2.2, the value of this threshold mass is 
very sensitive to the parameters of our model. The escape
fractions at masses $M_*\la 50M_\odot$ are not robust predictions of
our calculations, but are shown here for completeness.

\subsection{Ionization History} 

Shown in Figure \ref{mvt} is the evolution of the ionized gas mass
outside the halo, $M_{\rm HII}(t)$,  for different stellar masses.  
As expected, the more massive
the star, the more gas is ionized.  When expressed in units of the
mass of the star, however, the quantity 
$\eta_{\rm HII}\equiv M_{\rm HII}(t_*)/M_*$ is approximately constant 
with stellar mass for $M_*\ga 80$, $\eta_{\rm HII}\simeq 50,000-60,000$ 
(Figure \ref{etaII}).  In the
absence of recombinations, so that every ionizing photon results in
one ionized atom at the end of the star's lifetime, $\eta_{\rm
HII}=\eta_{\rm ph}$, where 
\beq 
\eta_{\rm ph}\equiv \frac{Q_*t_*m_p}{XM_*}
\eeq 
is the number of ionizing photons produced per stellar H atom
over the star's lifetime.  This efficiency is roughly independent of
mass for massive primordial stars $M_* \ga 50 M_\odot$, $\eta_{\rm ph}
\simeq 90,000-100,000$ for $X=0.75$ (e.g., Bromm et al. 2001;
Venkatesan, Tumlinson, \& Shull 2003; Yoshida, Bromm, \& Hernquist
2004).  Recombinations cause the value of $\eta_{\rm HII}$ to be lower than
$\eta_{\rm ph}$ by about a factor of two for large masses.

\subsection{IMF dependence}

Given the strong negative radiative feedback from H$_2$ dissociating
radiation that  is expected once a star forms within a minihalo (e.g.,
Haiman, Rees, \& Loeb 1997; Haiman, Abel \& Rees  2000), 
it is unlikely that more than one star
will exist there at any given time.  This negative feedback may extend
to nearby halos, though there is some uncertainty as to how strong
this negative feedback is (e.g., Ferrara 1998; Riccotti, Gnedin \& Shull
2002). Thus, the  first generation of stars forming within minihalos
likely formed in isolation, and  the initial mass function (IMF) of these stars was
probably determined by various properties of the host halos, such as
their angular momentum and accretion rate.  We make the reasonable
assumption that the  density structure of halos with a mass $M\sim
10^6 M_\odot$ is universal, so  that our determination of the escape
fraction for this halo is close to what  would be expected for other
halos of comparable mass.  For host halos of this  mass, therefore,
variations in the escape fraction come only from variations  in
stellar mass.  Under these assumptions, we can convolve
our results for one halo with different IMFs to see how the average
escape fraction depends on the IMF. Usually, when applied to
present-day star formation, the IMF describes the actual distribution
of stellar masses in 
a cluster consisting of many members.
In the primordial minihalo case, however, where stars are predicted to form
in isolation, as single stars or at most as small multiples, the
``IMF'' would be more appropriately interpreted as a `single-draw'
probability distribution (Bromm \& Larson 2004). Our analysis here is
carried out in this latter sense.

For definiteness, we
use a Salpeter-like functional form, given by 
\beq 
\Phi (M)=\left\{
\begin{array}{ll}KM^{-1.35}, & M_{\rm min}<M<M_{\rm max} \\ 0, 
& {\rm otherwise}, \\
\end{array}
\right.   
\eeq 
and normalized so that  
\beq 
\int_0^\infty \Phi(M)d\ln M = 1.  
\eeq 
\begin{inlinefigure}
\resizebox{8cm}{!}{\includegraphics{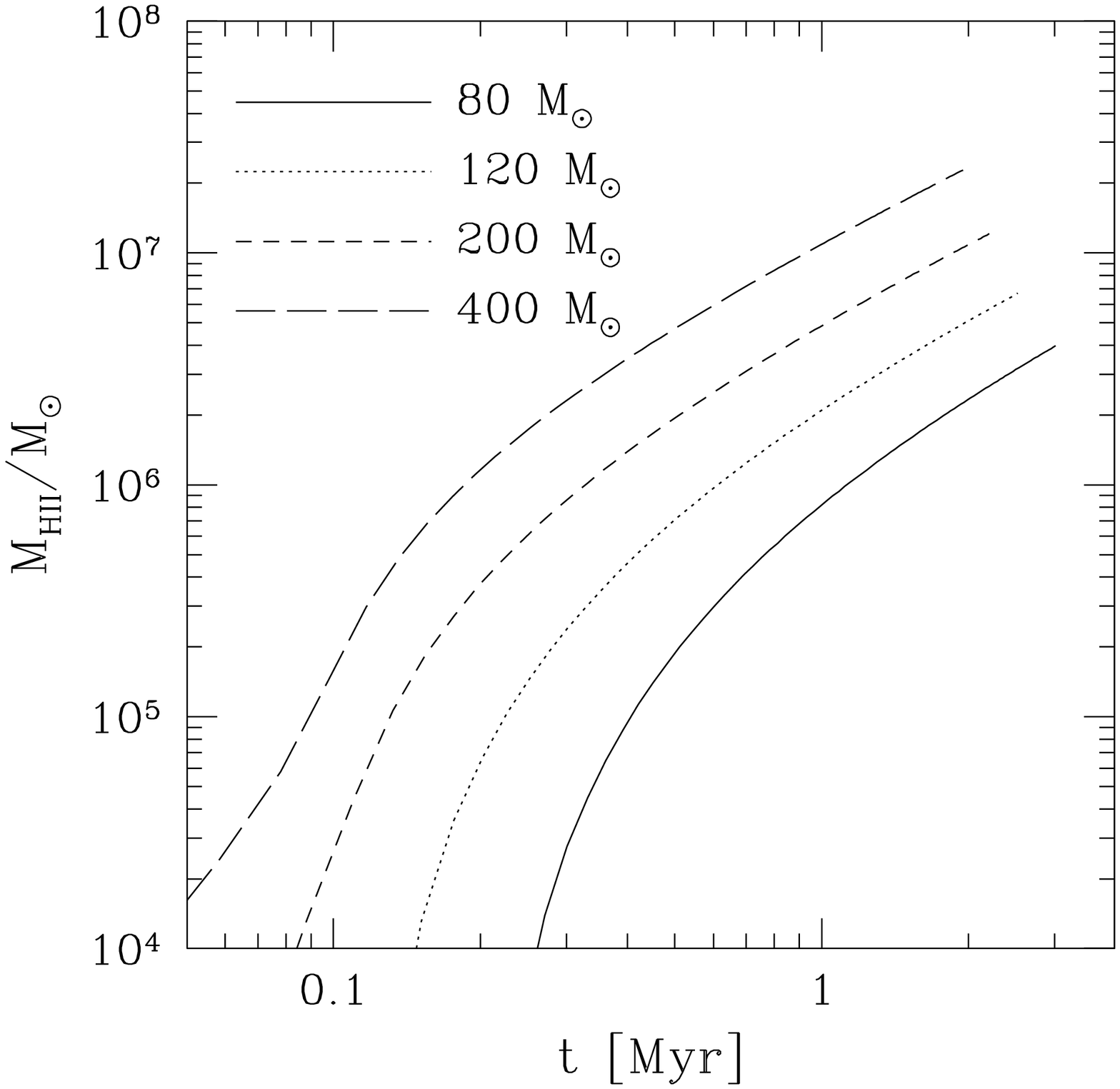}}
\caption{Mass ionized $M_{\rm HII}$ versus time for different stellar
masses, as labeled.  More massive stars produce more ionizing photons
in their lifetime and, therefore, ionize more of the surrounding gas.
\smallskip
\label{mvt}
}
\end{inlinefigure}
The total escape fraction, assuming one star forms per
halo of mass  $M\sim 10^6 M_\odot$, is given by  
\beq 
f_{\rm esc}^{\rm IMF}\equiv 
\frac {\int_0^\infty \Phi (M)Q_*(M)t_*(M)f_{\rm esc}(M)d\ln
M} {\int_0^\infty \Phi (M)Q_*(M)t_*(M)d\ln M}, 
\eeq 
where the total
number of photons released over a star's lifetime, $Q_*t_*$ appears in
the integrand because the escape fraction is being averaged over a
period of time which is long compared to the  lifetime of a star.
For $M_{\rm min}=0.5$ and
$M_{\rm max}=500$, which is a conservative estimate for the maximum
Pop~III stellar mass (Bromm \& Loeb 2004),
for example, the escape fraction is $\sim 0.5$,
whereas for $M_{\rm min}=0.5$ and $M_{\rm max}=80$, it is $\sim 0.35$.

\subsection{Structure of \ion{H}{2} region}
As can be seen from the visualization in Figure \ref{panels}, the
structure of the \ion{H}{2} region is highly asymmetric, with deep
shadows created by overdense gas.  In particular, nearby halos are not 
ionized, but rather are
able to shield themselves and all that is behind them from the
ionizing radiation of the star.  This can clearly be seen in the
bottom panels of Figure \ref{panels}, where overdense gas near to the
central star remains neutral, despite being so close.  Figure \ref{sph} 
shows the location of neutral and ionized SPH particles close to the
star, showing that the highest density gas nearby the star remains neutral. 
For example, the highest density of hydrogen that is ionized within
500 pc of the $120 M_\odot$
star is $\sim  2$ cm$^{-3}$ at a radius of $\simeq 200$ pc, which
corresponds to an overdensity  of $\delta \sim 4 \times 10^3$, whereas
the highest density of neutral hydrogen  is $\sim 400$ cm$^{-3}$ at a
radius of $\simeq 220$ pc, corresponding to $\delta  \sim 2.5 \times
10^5$.  Similarly overdense gas that is further from the star is even
more likely to shield itself and remain neutral, since the flux there
is weaker because of spherical  dilution.  
We find that $\sim 4.9\%$
of the sky at the end of the life of the $80 M_\odot$ star is covered
by high density gas that traps the I-front, whereas $\sim
2.6\%$ of the sky is covered at the end of the $200 M_\odot$ star's
life.  Such shielded regions are likely to be the sites of
photoevaporation (Shapiro, Iliev \& Raga 2004).  The photoevaporation
time of a $2\times 10^5 M_\odot$ halo which is at a distance of $250$
pc from a $120 M_\odot$ star with luminosity  $1.4\times 10^{50} {\rm
s}^{-1}$ is $\simeq 16$ Myr (Iliev, Shapiro, \& Raga 2005),  longer
than the lifetime of the star, so that most minihalo gas is likely to
retain its original density structure.

An important quantity associated with the ``relic \ion{H}{2}
region'' is the clumping factor.  Shown in Figure \ref{cvm} is the
clumping factor of the relic \ion{H}{2} region, $c_l\equiv 
\langle n^2\rangle/{\overline{n}}^2$, where the average is over the
volume of the \ion{H}{2} region and ${\overline{n}}$ is the cosmic
mean density.  Thus, the recombination time in the \ion{H}{2} region
is given by $t_{\rm rec}=t_{\rm rec,0}/c_l$, where $t_{\rm
rec,0}\simeq 100$ Myr is the recombination time of gas at the cosmic
mean density. As the  mass and 
luminosity of the star increase,
the clumping factor of the relic \ion{H}{2} region decreases.
Apparently, clustering of matter around the host halo causes the
clumping factor to increase near the halo.  Lower luminosity sources
leave behind smaller,  and therefore more clumpy, relic \ion{H}{2}
regions.  The mean recombination time in the regions, however, is
always less than the Hubble time $\sim 175$ Myr for even the largest
stellar masses, with recombination times $t_{\rm rec}< 60$ Myr.  These
\ion{H}{2} regions are thus likely to recombine unless other sources
are able to keep them ionized.   The timing of this  recombination and
the associated cooling of the recombining gas is crucial to
understanding the effect of photoheating on suppressing 
subsequent halo formation, the so-called ``entropy-floor'' 
(Oh \& Haiman 2003).

\subsection{I-front trapping by neighboring halos}
Whether or not nearby halos trap the I-front should determine whether
ionization stimulates star formation in their centers.   We can
estimate the radius and density at which trapping occurs, as follows.
The condition for trapping is  
\begin{inlinefigure}
\resizebox{8cm}{!}{\includegraphics{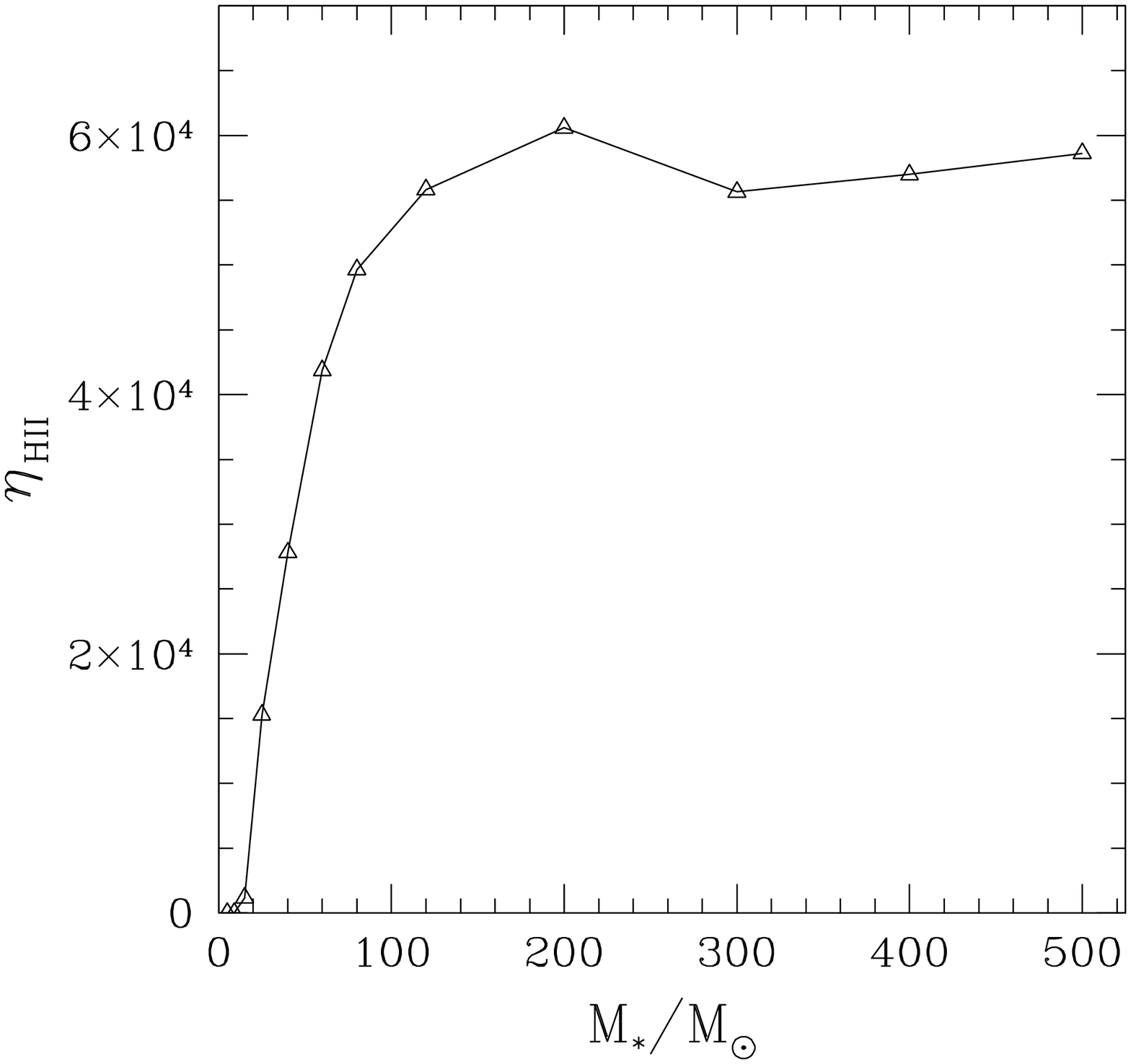}}
\caption{Ratio of ionized gas mass to stellar mass,
$\eta_{\rm HII}$, versus stellar mass. Notice that for
$M_* \ga 80 M_{\odot}$, this ratio is almost independent
of stellar mass.
\smallskip
\label{etaII}
}
\end{inlinefigure}
\begin{figure*}
{\includegraphics[width=\textwidth]{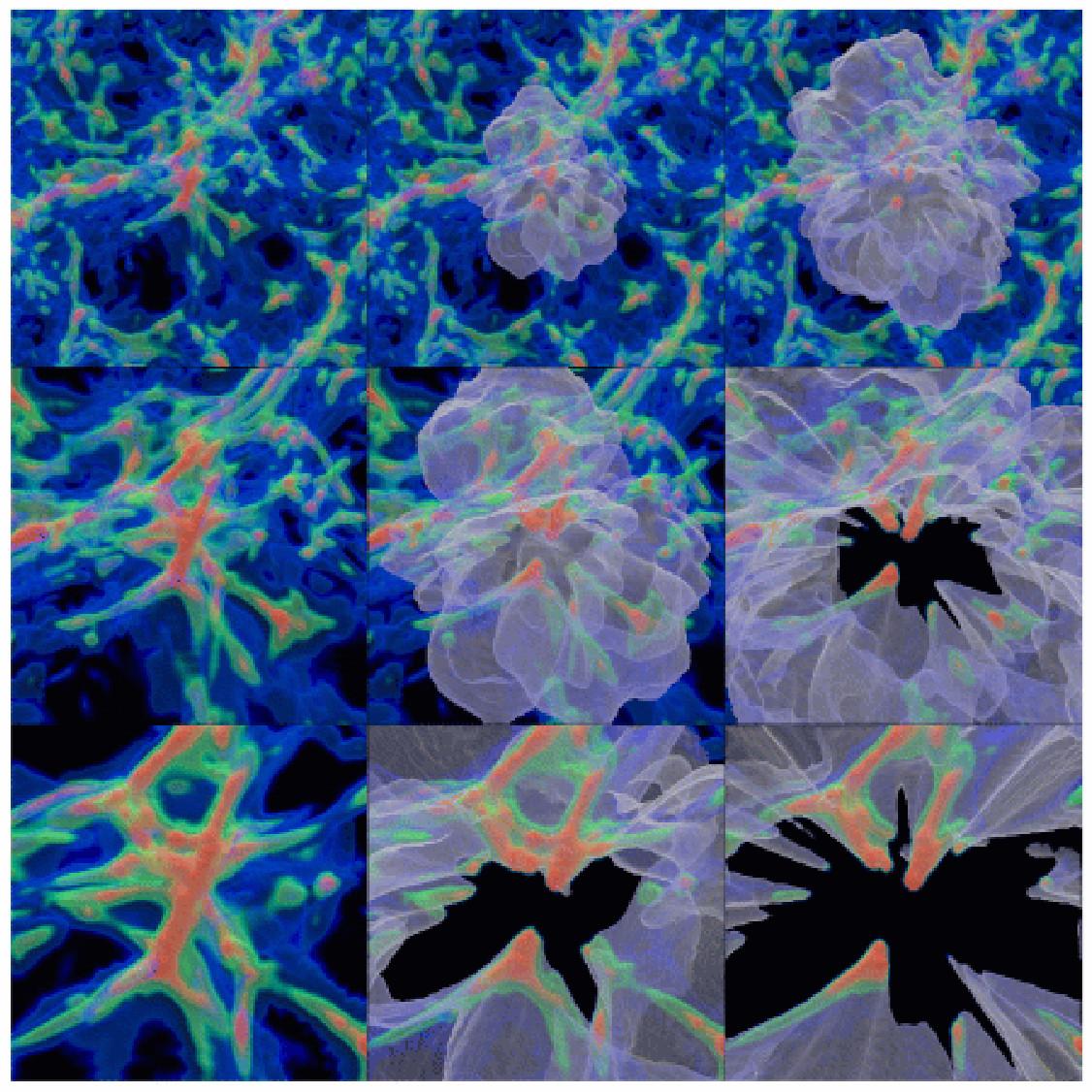}}
\caption{Volume visualization at $z=20$ of neutral density field 
(blue -- low
density,  red -- high density) and I-front (translucent white
surface).  Top row panels show a cubic volume $\sim 13.6$ kpc (proper) 
across, middle row $\sim 6.8$ kpc,  and bottom row $\sim 3.4$ kpc.  Left
column is at the initial time, middle column shows simulation at
$t_*=3$ Myr for the run with stellar mass  $M_*=80 M_\odot$, and the
right column shows simulation at $t_*=2.2$ Myr for the run with
stellar mass $M_*=200 M_\odot$.  The empty black region in the  lower
panels of middle and right columns indicates fully ionized gas around
the source, and is fully revealed as the volume visualized shrinks to
exclude the I-front that obscures this region in the larger
volumes above.
\label{panels}
}
\end{figure*}
\newpage
\beq  F=\int_{r_t}^{r_{\rm
vir}}\alpha_Bn_{\rm H}^2(r)dr,  
\eeq  
where $F$ is the external flux
of ionizing photons, $r_t$ is the radius at which the I-front
is trapped, and $r_{\rm vir}$ is the virial radius of the halo.  If we
assume that the halo has a singular isothermal sphere density profile
$n_{\rm H}(r)\propto r^{-2}$ and an overdensity $\delta_{\rm vir}$,
then solving for $r_t$ we obtain  \beq  \frac{r_t}{r_{\rm
vir}}=\left[1+\frac{9(36\pi)^{1/3} Fm_{\rm H}^2\Omega_m^2}{\alpha_BX^2
M_{\rm vir}^{1/3} \left( {\overline{\rho}(z)}\delta_{\rm
vir}\right)^{5/3}\Omega_b^2}\right]^{-1/3},  \eeq  where
${\overline{\rho}(z)}$ is the mean matter density of the universe at
redshift $z$.  For $r_t^3/r^3_{\rm vir}\ll1$, we can neglect the first
term in the brackets, to see how this trapping condition depends upon
the source and halo parameters and the redshift,  \beq
\frac{r_t}{r_{\rm vir}}\approx\left[
\frac{4\pi\alpha_B\rho_0^{5/3}X^2\Omega_b^2}{9(36\pi)^{1/3}m_{\rm
H}^2\Omega_m^2} \right]^{1/3} M_{\rm vir}^{1/9}\delta_{\rm
vir}^{5/9}Q_*^{-1/2}r^{2/3}(1+z)^{5/3},  \eeq  where $\rho_0$ is the
mean matter density at present.  The density at the point where the
I-front is trapped is  \beq  n_t\equiv n_{\rm H}(r_t) =
\frac{X{\overline{\rho}}(z) \Omega_b\delta_{\rm vir}} {3m_{\rm
H}\Omega_m}\frac{r^2_{\rm vir}}{r^2_t}.   \eeq  If we use the
parameters of the minihalo nearest to our source halo for fiducial
values,
\begin{eqnarray}
\frac{r_t}{r_{\rm vir}}&\approx & 0.18  \left(\frac{M_{\rm
vir}}{2\times 10^5 M_\odot}\right)^{1/9} \left(\frac{r}{220\ {\rm
pc}}\right)^{2/3} \left(\frac{\delta_{\rm vir}}{200}\right)^{5/9}
\nonumber \\ &\times &  \left(\frac{Q_*}{1.4\times 10^{50}\ {\rm
s}^{-1}}\right)^{-1/3} \left(\frac{1+z}{21}\right)^{5/3}
\end{eqnarray}
and
\begin{eqnarray}
n_t &\approx & 3.6\ {\rm cm}^{-3} \left(\frac{M_{\rm vir}}{2\times
10^5 M_\odot}\right)^{-2/9} \left(\frac{r}{220\ {\rm
pc}}\right)^{-4/3}\\ &\times &  \left(\frac{\delta_{\rm
vir}}{200}\right)^{-1/9}  \left(\frac{Q_*}{1.4\times 10^{50}\ {\rm
s}^{-1}}\right)^{2/3} \left(\frac{1+z}{21}\right)^{-1/3}\nonumber,
\end{eqnarray}
corresponding to a halo with total mass $M_{\rm vir}=2\times 10^5
M_\odot$ that is exposed to a flux $F=Q_*/(4\pi r^2)$ from a source
with luminosity $Q_*=1.4\times 10^{50} s^{-1}$ (for a stellar mass
$M_*=120 M_\odot$) at a distance $r=220$ pc.  Since $M(<R)\propto R$
in a singular isothermal sphere, $M_{\rm HI}/M_{\rm vir}=r_t/r_{\rm
vir}$, and thus the neutral gas mass for the fiducial case above is
$\simeq 5.5\times 10^3 M_\odot$.  The density at which the I-front is
trapped is much smaller than the central density expected for a
truncated isothermal sphere (Shapiro, Iliev, \& Raga 1999),  $n_{\rm
H,0}\simeq 30$ cm$^{-3}$ at $z=20$. Thus, nearby halos trap the
I-front well before it reaches the central core.  The weak dependence
of $r_t/r_{\rm vir}$ on halo mass and luminosity implies that trapping
is a generic occurrence for halos surrounding single primordial stars.

\section{Discussion}

We have studied the evolution of the \ion{H}{2} region created by a
massive Pop III star which forms in the current, standard $\Lambda$CDM
universe in a minihalo of total mass $M\sim 10^6 M_\odot$ at a
redshift of $z=20$.  We have performed a three-dimensional ray-tracing
calculation which tracks the position of the expanding I-front in
every direction around the source in the pre-computed density field
which results from a cosmological gas and N-body dynamics simulation
based on the GADGET tree-SPH code.  During the short lifetime
($\lesssim$ few Myr) of such a star, the hydrodynamical back-reaction
of the gas is relatively small as long as the front is a  supersonic
R-type, 
\begin{inlinefigure}
\resizebox{8cm}{!}{\includegraphics{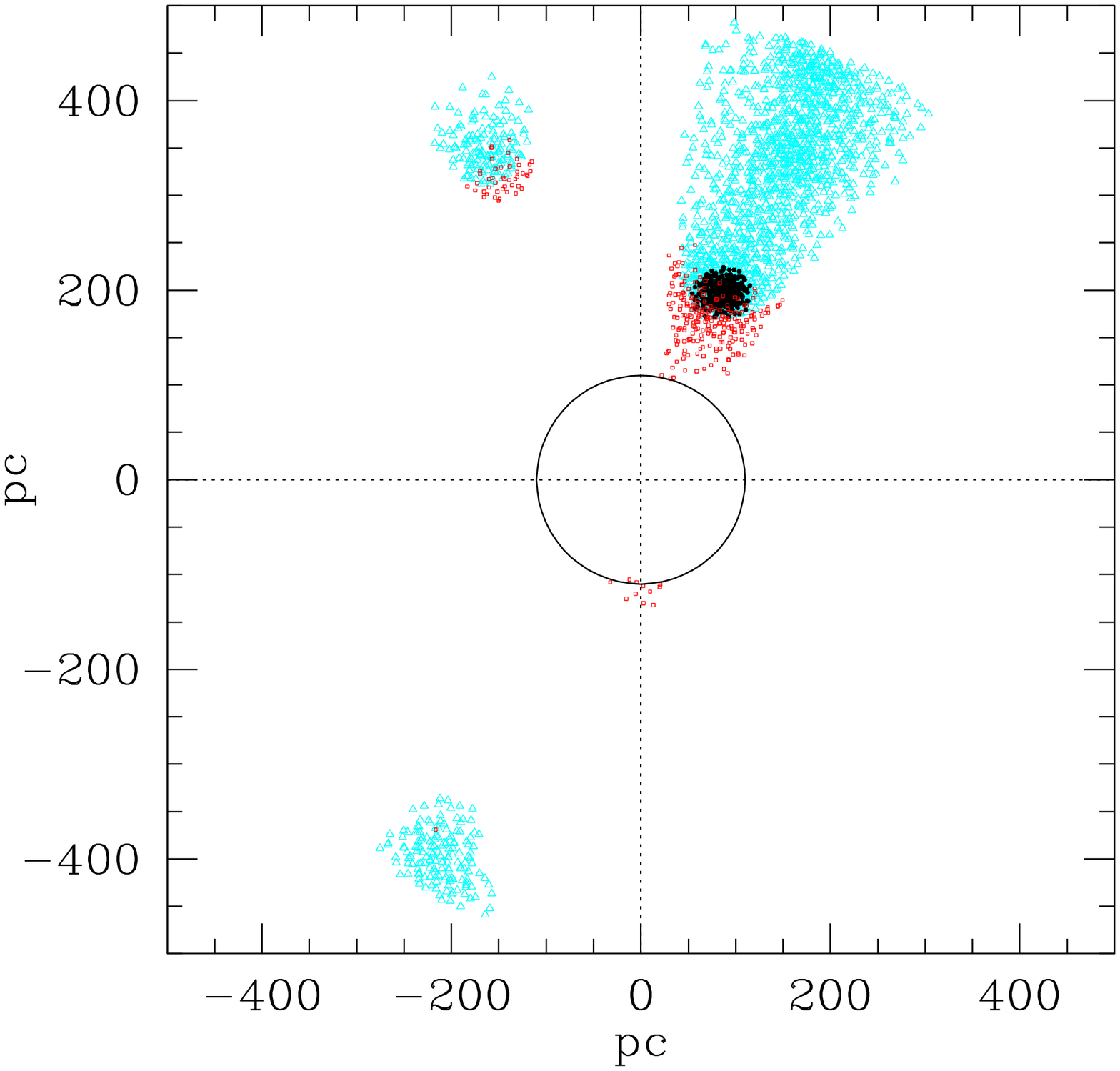}}
\caption{Position of selected SPH particles within 500 pc of the 120
$M_\odot$ star.
Red particles are ionized and have a density  above 1 cm$^{-3}$, all
the neutral particles are shown in cyan, while only neutral particles
with a density above 4 cm$^{-3}$ are colored black.  The radius of the
shock in the Shu solution at the end of the star's lifetime, $\sim 100$
pc, is shown as the circle in the center.  No SPH particles are shown
within that radius.
}
\smallskip
\label{sph}
\end{inlinefigure}
and, to first approximation, we are justified in  treating the
gas in this ``static limit.''  At early times, however, when  the
I-front is still deep inside the minihalo which formed the star, the
I-front is expected to make a transition from supersonic R-type to
subsonic D-type, preceded by a shock, before it eventually accelerates
to R-type again and detaches from the shock, racing ahead of it. 

To account for the impact of the expansion of the gas which results from
this dynamical phase on the propagation of the I-front after it
``breaks out,'' we  have used the similarity solution of Shu et
al. (2002) for champagne flow.  This solution allows us to determine
when the transition from D-type to R-type and ``break-out'' occurs
and, thereafter, to account for the consumption of ionizing photons in
the expanding wind left  behind in the central part of the minihalo.
In this way, we are able to track the progress of the I-front inside
the host minihalo and beyond, as it sweeps outward through the
surrounding IGM and encounters other minihalos.  This has allowed us
to investigate the link, for the first time, between the formation of
the first stars and the beginning of cosmic reionization on scales
close to the stellar source that could not be resolved in previous
three-dimensional studies of cosmic reionization.
Among the results of this calculation are the following.

Our simulations allow us to quantify the ionizing efficiency of the
first-generation of Pop~III stars in the $\Lambda$CDM universe as a
function of stellar mass.  The fraction of their ionizing radiation
which escapes from their parent minihalo increases with stellar mass.
For stars in the mass range $80\la M_*/M_\odot \la 500$, we find
$0.7\la f_{\rm esc}\la 0.9$.  This high escape fraction for high-mass
stars is roughly consistent with the high escape fraction found for
such high-mass stars by one-dimensional, spherical, hydrodynamical
calculations (Whalen et al. 
\begin{inlinefigure}
\resizebox{8cm}{!}{\includegraphics{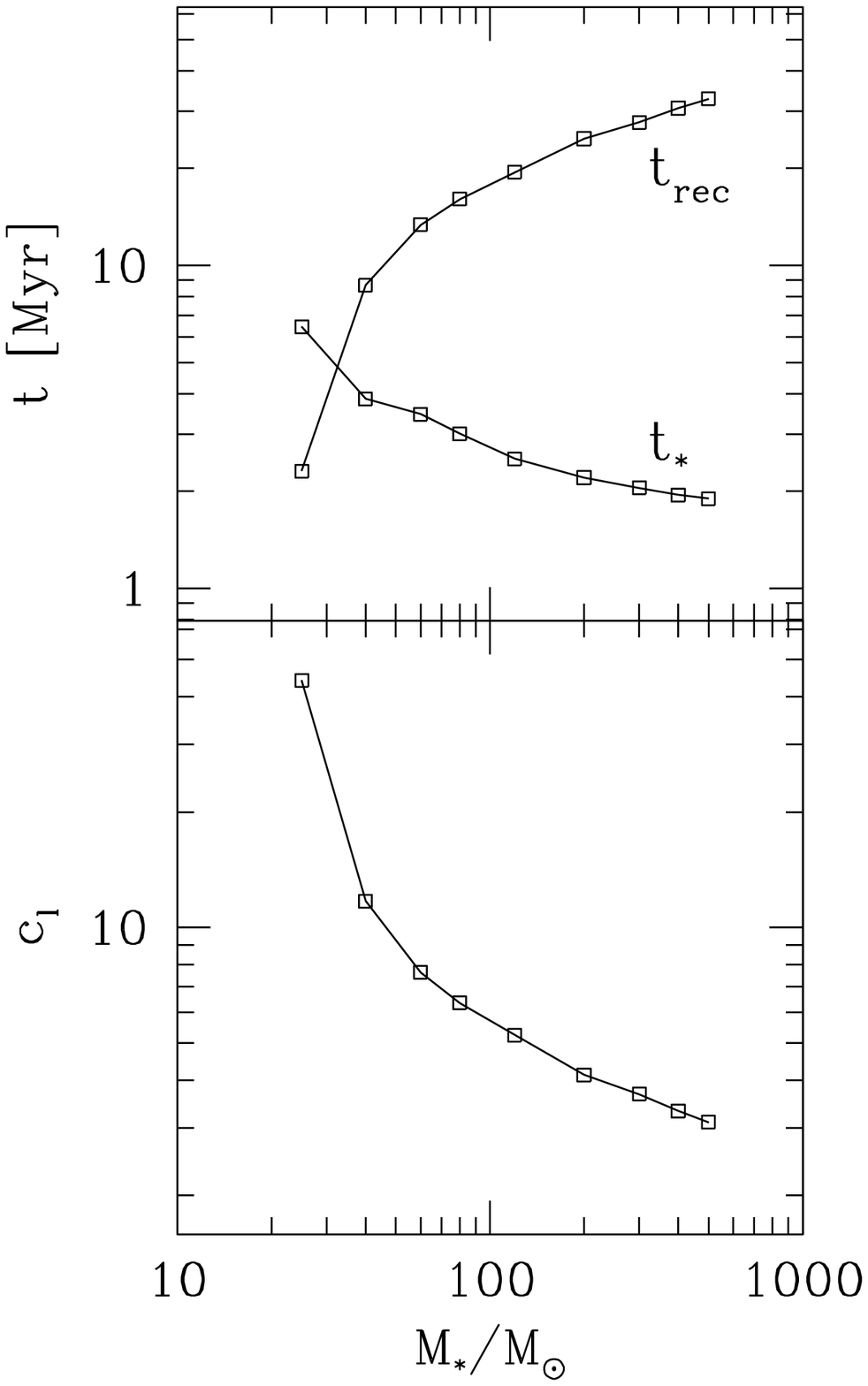}}
\caption{{\it Top:} Mean recombination time of \ion{H}{2} region at end of
star's life vs. stellar mass. {\it Bottom:} Clumping factor of \ion{H}{2}
regions at end of star's life vs. stellar mass.  Less massive stars
ionize a smaller volume, which implies a higher clumping factor because
of clustering around the host halo.
\smallskip
\label{cvm}
}
\end{inlinefigure}
2004; Kitayama et
al. 2004).  For lower-mass stars, the escape fraction drops more
rapidly with decreasing mass, as it takes a longer and longer fraction
of the stellar lifetime for the I-front to end the D-type phase by
reaching the ``break-out'' point, detaching from the shock and running
ahead as a weak, R-type front to exit the halo. For $M_*\la 15-20
M_\odot$, in fact, we find that the escape fraction
should be zero and the I-front is D-type for the whole life
of the star.  More importantly, we find that this threshold mass is
very sensitive to the hydrodynamic evolution of the I-front in the
D-type phase.  Given the great uncertainty regarding the interaction
of the stellar radiation and the gas immediately surrounding the star, a
definitive answer to this question can only be obtained through 
three-dimensional gas dynamical simulations with radiative
transfer that properly resolve the accretion flow onto the star.

Once the \ion{H}{2} region escapes the confines of the parent
minihalo, the reionization of the universe begins.  Our simulations
yield the ratio of the final total mass ionized by each of these
first-generation Pop~III stars to their stellar mass.  We find that,
for $M_*\ga 80M_\odot$, this ratio is about 60,000, roughly half the
number of H ionizing photons released per stellar atom during the
lifetime of these stars, independent of stellar mass.

We can obtain a very rough estimate of how effective Pop~III stars are
at reionizing  the universe by assuming that all the stars have the
same mass and form at the same redshift, each in its own minihalo of
mass $\sim 10^6 M_\odot$.  In the limit where the \ion{H}{2} regions
of individual stars are not overlapping, the ionized mass fraction fraction
is given in terms of the halo mass function by  $f_{\rm HII}=\rho_{\rm
HII}/{\overline{\rho_{\rm H}}}$ where $\rho_{\rm HII}\equiv \eta_{\rm
HII}M_*dn/d\ln M$ is the mean density of ionized gas, assuming each
halo hosts a star of mass $M_*$, and ionized a mass $\eta_{\rm HII}$
times the star's own mass, and ${\overline{\rho_{\rm H}}}$ is the
mean mass density of hydrogen (For $M_*\simeq 80M_\odot$, for example,
$\eta_{\rm HII}\simeq 50,000$).  Using the mass function of Sheth \&
Tormen (2002), $dn/d\ln M\simeq 130$ Mpc$^{-3}$ in comoving units at
$z=20$ for this mass range, we obtain $f_{\rm HII}\simeq 0.1
[\eta_{\rm HII}/5\times 10^4][M_*/80M_\odot]$.  If, instead, we use
the ionized volume $V_{\rm HII}$ obtained directly from our
calculations to derive the volume filling factor $f_{\rm V,HII}\equiv 
V_{\rm HII}dn/d\ln M$ for an $80 M_\odot$ star, the final ionized
volume is $7\times 10^{-4}$ comoving Mpc$^3$, corresponding to a
filling factor of $f_{\rm V,HII}(M_*=80M_\odot)\simeq 0.1$.  The
similarity of the volume and mass ionized fraction indicates that the
mean density in the ionized region is approximately equal to the mean
density of the universe.  

The above estimate is only the instantaneous ionized fraction, since
the recombination times of each relic \ion{H}{2}
region are small fractions of the age of the universe at $z=20$.  This
implies that many new generations of similar stars would have to form
to continuously maintain this ionized fraction.
A more conservative estimate of the effect of Pop~III stars on
reionization would also have to take account of the
back-reaction of the starlight from earlier generations of 
stars on the star formation rate in the minihalos
that follow (e.g., Mackey, Bromm, \& Hernquist 2003; Yoshida et al. 2003;
Furlanetto \& Loeb 2005).
Since Pop~III star formation depends upon the efficiency of H$_2$ formation
and cooling inside minihalos, a background of UV starlight between 11.2
eV and 13.6 eV is capable of suppressing this star formation by
photodissociation the H$_2$ following absorption in the Lyman-Werner
bands.  It is quite possible that the photodissociating background
builds up fast enough that minihalos are ``sterilized'' against
further star formation before such a large fraction of the universe
can be reionized in this way (Haiman et al. 1997).

Hydrodynamic feedback due to photoionization heating of the host halo
will have a dramatic impact on its ultimate fate.  Massive Pop~III
stars are expected to end their  lives either by collapsing to black
holes or exploding as pair-instability supernovae
(e.g., Madau \& Rees 2001; Heger et al. 2003).
In both cases, it
is important to know the properties of the host halo gas.  Our model for
the hydrodynamic feedback, in which an ionized, nearly-uniform density
bubble bounded by an isothermal shock propagates outward at a few
times the sound speed allows for an estimate of the density and size
of the bubble at the end of the star's lifetime.  For a lifetime of
2.5 Myr, the final size and density of the bubble are $r_{\rm
bubble}\simeq 100$ pc and $n_{\rm bubble}\la 1$ cm$^{-3}$.  If the
star ended its life by exploding, rather than collapsing to a black
hole, then the SN remnant evolution inside this low-density bubble and
beyond will differ from its evolution in the original undisturbed
minihalo gas.  This should be taken into account in models of the
impact of first-generation SN remnant blast-waves on their
surroundings (e.g., Bromm, Yoshida, \& Hernquist 2003).

This density can be used to estimate the accretion rate onto a possible
remnant black hole, $\dot{M}_B\simeq 4\pi G^2 M_{\rm BH}^2\rho/c_s^3$
(e.g., Bondi \& Hoyle 1944; Springel, Di Matteo, \& Hernquist 2005).
Thus, e.g., for a black hole mass of $100 M_\odot$ and a sound speed of 15
km~s$^{-1}$, we obtain  $\dot{M}_B\sim 2\times 10^{-5} M_\odot$
Myr$^{-1}$.  If we make the optimistic assumption that after a
recombination time ($\sim 1.2\times 10^{5}$ yr) the gas can form ${\rm
H}_2$ molecules and cool back to $\sim 300$ K without escaping from
the halo, the accretion rate increases by a factor of $10^3$, to
$\sim 2\times 10^{-2} M_\odot$Myr$^{-1}$.  These accretion rates are
small compared to  the Eddington accretion rate $\dot{M}_{\rm Edd}=4\pi GM_{\rm
BH}m_p/(\epsilon\sigma_T c)\simeq 2 M_\odot$ Myr$^{-1}$, where the
efficiency factor $\epsilon = 0.1$ (see also O'Shea et al. 2005). Such
low accretion rates imply 
that remnant black holes from the first generation of stars are
unlikely to be strong sources of radiation.  Calculations which do not
explicitly take into account this reduction in gas density near the 
remnant (e.g., Kuhlen \& Madau 2005) risk substantially
overestimating their radiative feedback as miniquasars.
These remnant black holes could begin to emit a substantial
amount of radiation only after they encounter some other environment
containing cold, dense gas.  It is not clear when or whether these black
holes would ever encounter such an environment.
At the very least, therefore, there should be some delay between the
formation of the first generation of stars and the X-ray emission 
from their remnants, if such emission ever occurs.   

Determining the fate of recombining gas in the relic \ion{H}{2} regions
left behind by the first stars is crucial.  The contribution of these
relic \ion{H}{2} regions from the first-generation Pop~III stars to
the increasing ionized fraction of the universe during cosmic
reionization depends upon their recombination time.  Because of clustering
around the host halo, the clumping factor and recombination time
within the relic \ion{H}{2} region depends on the mass of the star;
higher stellar masses correspond to lower clumping factors and longer
recombination times.  The recombination time and clumping factor for a
$40 M_\odot$ star are about 10~Myr and 12, respectively, whereas  for
a $500 M_\odot$ star they are about 35~Myr and 3 (see Fig. 10). 

When ionized gas within the relic \ion{H}{2} region cools radiatively
and recombines, the nonequilibrium recombination lags the cooling,
which enhances the residual ionized fraction at 10$^4$ K, promoting
the formation of H$_2$ molecules which can cool the gas to $T\sim 100$
K and enhance gravitational instability (Shapiro \& Kang 1987).  
This corresponds to ``positive feedback'' if further star formation
results (e.g., Ferrara 1998; Riccotti, Gnedin, \& Shull 2001).

Recently, O'Shea et al. (2005) have addressed this issue of possible
second generation star formation in the relic \ion{H}{2} region of the
first Pop~III stars.  They report that the I-front of the first star will
fully ionize the neighboring minihalos and that, when the initial star
dies, the dense cores of these ionized neighbor minihalos will be
stimulated to form H$_2$ molecules, leading to the second generation of
Pop~III stars.  This assumes the initial star collapses to form a black
hole without exploding as a supernova.  We find, however, that the
neighboring minihalos are {\em not} fully ionized before the initial star
dies, since the I-front is trapped by these minihalos and converted to
D-type outside the core region, and the photoevaporation time for the
minihalo exceeds the lifetime of the ionizing star.  It remains to be
seen, therefore, if the enhanced H$_2$ formation which O'Shea et al.
(2005) found inside the nearest neighbor minihalo in their simulations of
the relic \ion{H}{2} region will occur in the presence of the I-front
trapping and self-shielding which we report.  More work will be required
to resolve this question.

Whether H$_2$ molecules form in abundance or not depends on the density 
of the recombining gas. As we discussed in \S4.4, the highest density
of gas in the relic \ion{H}{2} region which we found to be fully
ionized and, hence, capable of 
recombining to form  ${\rm H}_2$ molecules, is a few cm$^{-3}$.  A
rough estimate of the ${\rm H}_2$ molecule  formation time is given by
the recombination time of the gas, $t_{\rm rec}\simeq 10^5$ yr.  Even
if ${\rm H}_2$ molecules form, however, it is not certain whether this
will promote star formation in neighboring halos.  As we have shown,
the densest gas located  in the center of nearby halos is not ionized.
The gas that is ionized is not in the center, and is thus likely to
recombine while being ejected from the halo as part of a supersonic,
photoevaporative outflow (e.g., Shapiro, Iliev, \& Raga 2004).  Such
gas is less likely to be gravitationally unstable.  In future work, we
will investigate whether this gas is able to cool and form stars or
simply gets evaporated into the diffuse IGM.


\acknowledgments
MAA is grateful for the support of a Department of Energy Computational
Science Graduate Fellowship. VB acknowledges support from NASA {\it Swift}
grant NNG05GH54G. This work was partially supported by NASA Astrophysical
Theory Program grants NAG5-10825 and NNG04G177G and Texas Advanced
Research Program grant 3658-0624-1999 to PRS.  The cosmological
simulations used in this work were carried out at the Texas Advanced
Computing Center (TACC).


\appendix
\section{Mass-conserving SPH Interpolation onto a Mesh}
Rather than interpolate directly from the SPH particles to our
spherical grid of rays, we first interpolate the density to a uniform
rectilinear mesh.  The assignment  of density to the mesh should conserve mass,
which we accomplish as follows.  We use a Gaussian kernel 
\beq
W(r,h)=\frac{1}{\pi^{3/2}h^3}e^{-r^2/h^2}, 
\eeq 
where $h$ is the smoothing length and $r$ is distance.  This kernel is
very similar to the commonly-used spline kernel,
\begin{equation}
W(r,h) = \frac{8}{\pi h^3}\left\{
\begin{array}{ll} 
1-6\left(\frac{r}{h}\right)^2+6\left(\frac{r}{h}\right)^3, &
0\leq\frac{r}{h}\leq \frac{1}{2},\\ 2\left(1-\frac{r}{h}\right)^3, &
\frac{1}{2}\leq \frac{r}{h}\leq 1,\\ 0 & \frac{r}{h}>1.
\end{array}  
\right.
\end{equation}
In our case, where a spline kernel has been used in the SPH
calculation, we find that a Gaussian  kernel is sufficient for
interpolation purposes, provided we transform the smoothing lengths
according to $h_{\rm Gauss}=\pi^{-1/6}h_{\rm spline}/2$.

For interpolation to a uniform rectilinear mesh with cell size
$\Delta_c$, we wish to find the mean density within a cell centered
at $(x,y,z)$ contributed by a particle with smoothing length $h$
located at the origin, 
\beq 
{\overline W}({\bf
r},h)\equiv\frac{1}{\Delta_c^3}\int_V W(r,h), 
\eeq 
where the integral
is over the volume of the cell.  Since the kernel is a Gaussian, the
spatial  integral separates into three separate ones, so that 
\beq
{\overline W}({\bf r},h)=\frac{1}{8\Delta_c^3}\Xi(x)\Xi(y)\Xi(z),
\eeq where 
\beq 
\Xi(s)\equiv{\rm
erf}\left[\frac{s+\Delta_c/2}{h}\right]- {\rm
erf}\left[\frac{s-\Delta_c/2}{h}\right].  
\eeq 
The value of the density
averaged over a cell centered at ${\bf r}_c$ is thus 
\beq 
{\overline\rho}({\bf r}_c)=\sum_i m_i {\overline W}({\bf r}_c-{\bf r}_i,h_i),
\eeq 
where the sum is over all particles $i$ such that  ${\overline
W}({\bf r}_c-{\bf r}_i,h_i)/{\overline W}(0,h_i)>\epsilon$, so as not
to needlessly sum over particles with a negligible contribution.  We
find that a value $\epsilon=10^{-5}$ is sufficient  for our purposes.


\begin{thebibliography}{11}

\bibitem{ABN} 
Abel, T., Bryan, G. L., \& Norman, M. L. 
2002, Science, 295, 93

\bibitem{BL01}
Barkana, R., \& Loeb, A. 2001, Phys. Rep., 349, 125

\bibitem{BAC84}
Bond, J. R., Arnett, W. D., \& Carr, B J. 1984, ApJ, 280, 825

\bibitem{BH}
Bondi, H., \& Hoyle, F. 1944, MNRAS, 104, 273

\bibitem{Br99} 
Bromm, V., Coppi, P. S., \& Larson, R. B. 
1999, ApJ, 527, L5 

\bibitem{Br} 
Bromm, V., Coppi, P. S., \& Larson, R. B. 
2002, ApJ, 564, 23 

\bibitem{BKL} 
Bromm, V., Kudritzki, R. P., \& Loeb, A. 2001, ApJ, 552, 464

\bibitem{BLrev} 
Bromm, V., \& Larson, R. B. 
2004, \araa, 42, 79

\bibitem{BLoeb03} 
Bromm, V., \& Loeb, A. 
2003, Nature, 425, 812

\bibitem{BLoeb} 
Bromm, V., \& Loeb, A. 
2004, NewA, 9, 353

\bibitem{PaperI} 
Bromm, V., Yoshida, N., \& Hernquist, L. 
2003, 
ApJ, 596, L135

\bibitem{cen03} 
Cen, R. 2003, ApJ, 591, 12

\bibitem{CF05}
Ciardi, B., \& Ferrara, A. 2005, Space Sci. Rev., 116, 625

\bibitem{CFW}
Ciardi, B., Ferrara, A., \& White, S.D.M. 2003, MNRAS, 344, L7

\bibitem{Couchman} 
Couchman, H. M. P., \& Rees, M. J.
1986, MNRAS, 221, 53

\bibitem{fer98}
Ferrara, A. 1998, ApJ, 499, L17

\bibitem{Franco}
Franco, J., Tenorio-Tagle, G., \& Bondenheimer, P. 1990, ApJ, 349, 126

\bibitem{FurL05} 
Furlanetto, S. R., \& Loeb, A.
2005, ApJ, in press (astro-ph/0409656)

\bibitem{GO} 
Gnedin, N. Y., \& Ostriker, J. P.
1997, ApJ, 486, 581

\bibitem{Gorski}
G\'orski, K. M., Hivon E., Banday A. J., Wandelt, B. D., 
Hansen, F. K., Reinecke, M., \& Bartelmann, M. 2005 ApJ, 622, 759

\bibitem{HAR00}
Haiman, Z., Abel, T., \& Rees, M. J. 2000, ApJ, 534, 11

\bibitem{HH}
Haiman, Z. \& Holder, G. P. 2003, ApJ, 595, 1

\bibitem{hrl6}
Haiman, Z., Rees, M. J., \& Loeb, A.
1997, ApJ, 476, 458

\bibitem{HTL96}
Haiman, Z., Thoul, A. A., \& Loeb, A. 1996, ApJ, 464, 523

\bibitem{Heg03} 
Heger, A., Fryer, C. L., Woosley, S. E., Langer, N., \& Hartmann D. H.
2003, ApJ, 591, 288

\bibitem{ISR}
Iliev, I. T., Shapiro, P. R., \& Raga, A. C. 2005, MNRAS, 361, 405

\bibitem{Kitay}
Kitayama, T., Yoshida, N., Susa, H., \& Umemura, M. 2004, ApJ, 613, 631

\bibitem{KO}
Kogut, A., et al. 2003, ApJS, 148, 161

\bibitem{KM05}
Kuhlen, M., \& Madau, P. 2005, MNRAS, in press (astro-ph/0506712)

\bibitem{MB} 
Mackey, J., Bromm, V., \& Hernquist, L. 
2003, ApJ, 586, 1

\bibitem{MR} 
Madau, P., \& Rees, M. J.
2001, ApJ, 551, L27

\bibitem{NU} 
Nakamura, F., \& Umemura, M. 
2001, ApJ, 548, 19

\bibitem{OAWN}
O'Shea, B. W., Abel, T., Whalen, D., \& Norman, M. L. 2005, ApJ, 626, L5

\bibitem{OH}
Oh, P., \& Haiman, Z. 2003, MNRAS, 346, 456

\bibitem{OI}
Omukai, K., \& Inutsuka, S. 2002, MNRAS, 332, 59

\bibitem{OP01}
Omukai, K., \& Palla, F. 2001, ApJ, 561, L55

\bibitem{OP03}
Omukai, K., \& Palla, F. 2003, ApJ, 589, 677
 
\bibitem{RGS}
Ricotti, M., Gnedin, N. Y., \& Shull, M. J. 2001, ApJ, 560, 580

\bibitem{sch02} 
Schaerer, D. 2002, A\&A, 382, 28

\bibitem{SG}
Shapiro, P. R., \& Giroux, M. L. 1987, ApJ, 321, L107

\bibitem{SIAS}
Shapiro, P. R., Iliev, I. T., Alvarez, M. A., \& Scannapieco, E. 2005,
ApJ, submitted (astro-ph/0507677)

\bibitem{SIR99}
Shapiro, P. R., Iliev, I. T., \& Raga, A. C. 1999, MNRAS, 307, 203

\bibitem{SIR04}
Shapiro, P. R., Iliev, I. T., \& Raga, A. C. 2004, MNRAS, 348, 753

\bibitem{SK}
Shapiro, P. R., \& Kang, H. 1987, ApJ, 318, 32

\bibitem{ST}
Sheth, R. K., \& Tormen, G. 2002, MNRAS, 329, 61

\bibitem{Shusol}
Shu, F. H., Lizano S., Galli, D., Cant\'o, J., \& Laughlin, G. 2002,
ApJ, 580, 969

\bibitem{sok} 
Sokasian, A., Yoshida, N., Abel, T., Hernquist, L., \& Springel, V. 
2004, MNRAS, 350, 47

\bibitem{SPE}
Spergel, D. N., et al. 2003, ApJS, 148, 175

\bibitem{Spitzer}
Spitzer, L. 1978, Physical Processes in the Interstellar Medium (New
York: Wiley)

\bibitem{SDH}
Springel, V., Di Matteo, T., \& Hernquist, L. 2005, MNRAS, in press 
(astro-ph/0411108)

\bibitem{SYW01}
Springel, V., Yoshida, N., \& White, S. D. M. 2001, NewA, 6, 79

\bibitem{TM}
Tan, J. C., \& McKee, C. F. 2004, ApJ, 603, 383

\bibitem{tegmark} 
Tegmark, M., Silk, J., Rees, M. J., Blanchard, A., Abel, T.,
\& Palla, F.
1997, ApJ, 474, 1

\bibitem{jason} 
Venkatesan, A., Tumlinson, J., \& Shull, J. M.
2003, ApJ, 584, 621

\bibitem{WAN}
Whalen, D., Abel, T., \& Norman, M. 2004, ApJ, 610, 14

\bibitem{wl2} 
Wyithe, J. S. B., \& Loeb, A. 
2003, ApJ, 588, L69

\bibitem{yos} 
Yoshida, N., Abel, T., Hernquist, L., \& Sugiyama, N. 
2003a, ApJ, 592, 645

\bibitem{yBH04} 
Yoshida, N., Bromm, V., Hernquist, L.
2004, ApJ, 605, 579

\bibitem{Yu}
Yu, Q., 2005, ApJ, 623, 683

\end{thebibliography}
\end{document}